\documentclass{iopart}
\usepackage[latin1] {inputenc}
\usepackage{amsfonts}
\usepackage{amssymb}
\usepackage{graphicx}
\usepackage{eufrak}
\newcommand{\rd}{\mathrm{d}}
\newcommand{\uni}{\mathrm}

\begin{document}
\title[Decaying neutron propagation in the Galaxy]{Decaying neutron 
propagation in the Galaxy and \\
the Cosmic Ray anisotropy at 1 EeV}

\author{Mat\'ias Bossa, Silvia Mollerach and Esteban Roulet}

\address{Centro At\'omico Bariloche, Av. Bustillo 9500, 8400 Bariloche,
Argentina }

\begin{abstract}
We study the cosmic ray arrival distribution expected from a source of
neutrons in the galactic center at energies around 1~EeV and compare
it with the anisotropy detected by AGASA and SUGAR. Besides the point-like
signal in the source direction produced by the direct neutrons, an
extended signal due to the protons produced in neutron decays is
expected. This associated proton signal also leads to an excess in the
direction of the spiral arm.
For realistic models of the regular and random galactic
magnetic fields, the resulting anisotropy as a function of the energy is
obtained. We find that for the anisotropy to become sufficiently
suppressed  below $E\sim 10^{17.9}$eV, a significant random magnetic
field component is required, while on the other hand, this also tends
to increase the angular spread of the associated proton signal and to
reduce the excess in the spiral arm direction.
The source luminosity required in order that the right ascension
anisotropy be $4\%$ for the AGASA angular exposure corresponds to a
prediction for the point-like flux from direct neutrons compatible 
with the flux detected by SUGAR.
We also analyse the distinguishing features predicted for a large
statistics southern observatory.
\end{abstract}

\pacno{98.70.Sa}

\maketitle

\section{Introduction}         

The distribution of the arrival directions of high energy Cosmic Rays (CRs)
provides a useful tool to investigate their origin. This distribution
is found to be very close to isotropic up to the highest energies
measured. For energies below the knee ($E_{knee}\simeq 3\times
10^{15}$~eV) the measured anisotropy is below $0.1\%$, showing a
smooth increase for larger energies and approaching the percent level
around $10^{18}$~eV \cite{hi84}. 
This isotropy is understood as due to the diffusive motion of
charged particles in the galactic magnetic field for energies such that
$E/Z\leq 10^{17}$~eV, with $Z$ the CR charge.
Cosmic rays from galactic sources remain trapped in this field, leading
to a flux building up at small energies and to the absence of correlations
between the arrival direction of CRs and the source's direction. 
At energies above 
$10^{19}$~eV, and up to the highest observed energies, the distribution
of the events is compatible with isotropy. However, at these
energies CRs are no more trapped in the galactic magnetic field, so that a
correlation between the arrival and the source directions is to be
expected. The absence of a significant correlation between ultra high
energy events and the Galactic plane points then to extragalactic sources
as the origin of these events.
 
In 1999 the AGASA group reported the measurement of the most
significant anisotropy detected so far \cite{agasa}. They presented the
results of a search for anisotropies in the arrival directions of
cosmic rays with energy above $10^{17}$~eV, using
data collected over 11 years. They found a strong anisotropy
of amplitude $\sim 4\%$ in a small energy range: $10^{17.9}-10^{18.3}$~eV.
The phase of the first harmonic in right ascension was found to be 
$\alpha_1 \simeq 300^\circ$.
A two dimensional map showed that this anisotropy can be
interpreted as due to excesses of events in two regions of size 
$\sim 20^{\circ}$ near the galactic center (GC) and the Cygnus region 
with significance $4\sigma$ and $3\sigma$, respectively.
A subsequent analysis by the same group, including an independent data
set, confirmed the initial claim \cite{ha99}.

Taking into account that the excess reported by AGASA lies close to
the galactic center, that is better seen from the south, 
the SUGAR group reanalyzed their data \cite{sugar}. The SUGAR experiment,
located near Sidney, has a clear
view of the GC, but has much poorer statistics than AGASA.
They set a priori the energy range corresponding to the AGASA
anisotropy and found a point like excess at $7.5^\circ$ from the GC,
and $6^\circ$ away from  the AGASA maximum. Due to the difference in
direction and angular extension of the AGASA and SUGAR signals it is
not yet clear whether they are consistent among each other and if they
can result from a common origin. 

A possible explanation of the anisotropy, already proposed by
Hayashida et al. \cite{agasa}, is that it is due to neutron primary
particles. Neutrons of $10^{18}$~eV have a Lorentz factor $\gamma\simeq 
10^9$, so that
their decay length is about 10~kpc, i.e. similar to the distance from the
Earth to the galactic center. Hence, neutrons from a source near the
galactic center can propagate up to the Earth before decaying, and
they are undeflected by intervening magnetic fields, contrary to the
case of charged primaries. The lack of significant  anisotropies at
smaller energies is explained within this scenario
as due to the shorter decay length of
the neutrons, and at larger energies by a cutoff in the acceleration
energy (or due to a very steep spectrum) of the sources. 
They suggested that neutrons can be produced by
accelerated heavy nuclei which disintegrated due to interactions with ambient 
matter or photons surrounding the source. Charged particles could stay
confined by the magnetic fields present in the acceleration region, while 
neutrons would escape easily. 
 
Medina-Tanco and Watson \cite{tanco} proposed that a more likely way
of generating a high energy neutron flux is through high energy
protons via interactions  with ambient protons or with infra red
photons. They found 
that the neighborhood of the Galaxy's central supermassive
black hole could give the environment needed for
these reactions to happen frequently enough.
Nagataki and Takahashi showed that the proton-proton process was the most
efficient one, and also studied the detectability of the $\gamma$ and 
$\nu$
emission associated to the production of neutrons in the GC \cite{na01}. 

The possibility that the GC is a source of high energy protons was
studied by Levinson and Boldt \cite{Sgr A}. They showed that, under
certain conditions, a compact black hole dynamo associated with the
Sgr A* source may accelerate protons up to energies $\sim 10^{18}$ eV,
with a total luminosity of $\sim 10^{38}$ erg/s.

In this work, we study in detail the CR arrival direction
distribution expected from a source of neutrons in the GC.
The fact that the decay length of $10^{18}$~eV neutrons is similar
to the distance to the proposed source implies that a sizable fraction
of the neutrons will actually decay to protons before reaching the
Earth, and the trajectories of these secondary protons will be bent by
the galactic  magnetic fields (GMF). 
These protons would arrive to the Earth from some preferred directions in
the sky: those produced near the Earth would come from directions close to
the GC one, while those produced in the inner Galaxy region would arrive
preferentially from directions close to the spiral arm, as their
trajectories wind around the regular magnetic field lines. 
Initial work in this direction was performed by Medina-Tanco and Watson
\cite{tanco} and here we extend it to take into account different allowed
magnetic field strengths, we map the arrival direction distribution for
different energies and take into account the angular exposure of
experiments at different latitudes to study several features of the
observations such as the energy dependence of the anisotropies, the
compatibility of the observations of AGASA and SUGAR both regarding
the angular extent of the observed features and the required source
luminosity.  

Other suggested explanations for the observed anisotropies include
a diffusive flux of charged particles from the GC direction
\cite{te01}, the drift of heavy nuclei in the galactic
magnetic field \cite{ca02} and a source of protons in the galactic
center \cite{cl00,be02}.

\section{Cosmic Ray propagation in the Galaxy}

The trajectories of charged CRs will be determined by the galactic
magnetic field, which has both a large scale regular component and a
random component, both with strengths of a few
$\mu$G. The magnetic lines of the regular field follow the spiral
structure, reversing direction between consecutive arms. According to
the preferred model for our galaxy, it is symmetric with respect to
the galactic plane.  
The strength of the field can be modeled as \cite{stanev,regular field}
\begin{eqnarray}
    B_{sp}(\rho,\theta,z) = B_0\frac{r_0}{2\rho}\tanh^3
\left(\frac{\rho}{\rho_1}\right)\ 
    \cos(\theta - \beta \ \ln \frac{\rho}{\xi_0})\nonumber\\
\bigg(\frac{1}{\cosh(z/z_1)}+\frac{1}{\cosh(z/z_2)}\bigg),
\end{eqnarray}
where $(\rho,\theta,z)$ are cylindrical coordinates with origin in the
galactic center,
$\xi_0=10.55$ kpc is the value of $\rho$ for which the field
is maximum in our spiral arm, $\beta=1/\tan\ p =-5.67$,
where $p=-10^\circ$ is the pitch angle and
$r_0=8.5$ kpc is the distance from the Sun to the GC.
The overall intensity is taken as $B_0=3~\mu$G, corresponding to a local
strength of the regular component close to $2~\mu$G.
The $\tanh$ term smoothes the $1/\rho$ behavior of the field near the
center within a core radius taken as $\rho_1=2$ kpc. The vertical
scale length associated to the disk field is taken as $z_1=0.3$ kpc,
while the halo one as $z_2=4$ kpc. 

The radial and azimuthal components are given by $ B_\rho=B_{sp}\ 
\sin~p$ and $B_\theta=B_{sp}\ \cos~p$ respectively, while
the strength of the field in the $z$ direction is assumed to vanish. 

Besides the regular magnetic field structure, a random component due
to turbulence in the interstellar plasma is known to exist, with the
largest eddies having a scale of $L_{max} \simeq 100$ pc. A usual
assumption is that the spectrum of the field inhomogeneities is the
same as that of the gas density, that is close to a  Kolmogorov
spectrum. Hence, the turbulent magnetic field can be described by a
random field with a power law spectrum such that the magnetic energy
density satisfies $dE/dk \propto k^{-5/3}$.
It can then be written as a sum of Fourier modes 
\begin{equation} \label{superpos}
  B_i(\vec x) = \int \frac{d^3k}{(2\pi)^3}
  B_i(\vec k) e^{i(\vec k . \vec x +
  \phi_i(\vec k))},
\end{equation}
where the phases $\phi_i(\vec k)$ are random and the amplitude of the
modes is given by 
\begin{equation} \label{espectro}
  B^2(k) = \frac{2}{3}B^2_{rms} k^{-5/3} \frac{(2 \pi /L_{max})^{2/3}}
  {1-(L_{min}/L_{max})^{2/3}},
\end{equation}
for $2 \pi / L_{max} \leq k \leq 2 \pi/L_{min}$, and zero
otherwise \cite{turbulent field}.
$B_{rms}$ fixes the root mean square amplitude of the random
component.
 In the numerical calculations the integral in Equation~(\ref{superpos}) was
 replaced by a sum over 200 modes. 
Each $\vec B (\vec k_i)$ had a random amplitude extracted
 from a Gaussian distribution with zero mean and root mean square 
value given by the square root of Equation~(\ref {espectro}) and  
the direction was chosen randomly in the plane perpendicular 
to $\vec k_i$, ensuring that $ \vec \nabla \cdot \vec B (\vec x)=0 $. 
Since particles reaching the Earth at the energies under study never
 departed from the galactic plane by more than a few hundred pc, the
 detailed vertical profile of the random component is not relevant for
 the results, and hence a constant value for it was adopted
 throughout. Moreover, the random field was  also taken as having a
 constant rms strength in the radial direction. 

The effect of these field components on the motion of EeV protons
can be estimated as follows.
The gyroradius of a high energy CR ($v \approx c$) with
charge $Z e$ and energy $E$ traveling through an
uniform magnetic field of strength $B$ is
$R_{\uni{kpc}} \simeq 1.1 E_{\uni{EeV}}/Z B_{\mu \uni G}$. Hence, a
proton with $E = 1$ EeV in a $B \sim 3~\mu \uni G$ field has a
gyroradius $R \sim 0.4$ kpc. As the regular GMF is uniform over scales
of the order of a few kpc and the thickness of the galactic disc is
$\sim$ 1 kpc, the trajectory of an EeV proton is expected to wind
around the magnetic field lines.

The turbulent component adds random deflections to the propagation. The
{\it rms} deflection of a CR with energy $E$ propagating a distance
$L$ through a turbulent magnetic field of strength $B_{rms}$
and maximum inhomogeneity scale $L_{max}$ 
(for $L \gg L_{max} $) is given by \cite{turbulent field}
\begin{equation} \label{deflection}
  \delta_{rms}(L)
  \simeq 27^\circ \frac{\uni{EeV}}{E/Z} \frac{B_{rms}}{1~\mu\uni G}
  \sqrt{\frac{L}{10~\uni{kpc}}}\sqrt{\frac{L_{max}}{100~\uni{pc}}}.
\end{equation}
Thus, we expect that for EeV protons traveling in the galactic turbulent
 field, this will induce a significant dispersion in the
arrival directions.

\section{Results}

\subsection{Arrival direction distribution}
\label{add}

Neutrons emitted from the GC travel along straight lines
until the moment in which they decay into protons, and it is only the
trajectories of these last which start to be bent by the GMF. Thus, in 
order to obtain the
distribution of arrival directions at Earth we should follow the
trajectories of protons injected in spheres of various radius around
the GC and with velocities pointing out of the GC, and record the
direction of those arriving to the Earth. These would have to be
weighted by the probability of neutron decay at the corresponding
radius. However, a more efficient way to compute this distribution is
to back track antiprotons leaving the Earth in all different
directions and record the points where their velocities point toward
the GC. It is clear that if a neutron from the GC were to decay at
that point, the proton produced would arrive to the Earth. We actually
considered a source with finite radius, and obtained the distance from
the GC ($r$) and the length of the
segment ($\Delta r$) of the proton trajectories for which the velocity
pointed towards the source. The corresponding flux of protons from
neutron decays that arrive from a given direction is then 
${\rm d}J/{\rm d}\Omega \propto e^{-r/\gamma c \tau_n} 
\Delta r/\gamma c \tau_n$.
Adding finally the contribution from
neutrons which did not decay, the arrival distribution map can be constructed. 

\begin{figure} [!ht]
\begin{center}
\includegraphics[width=12cm]{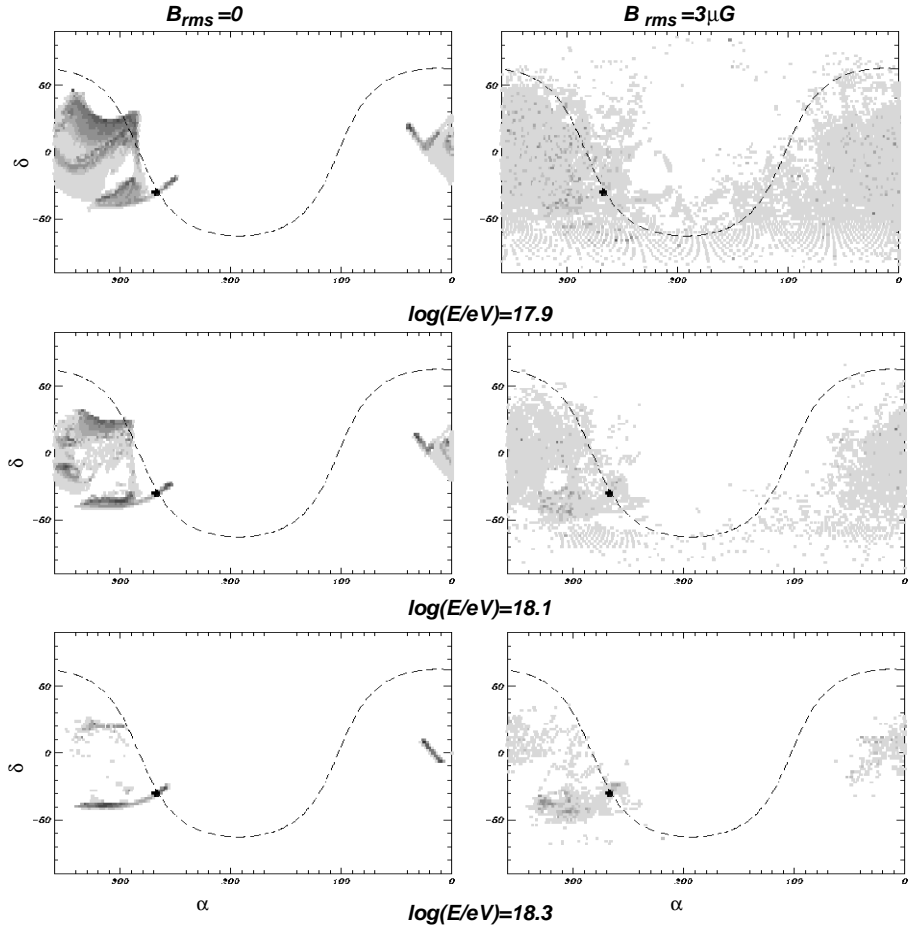}
\end{center}
\caption{Arrival direction distribution in celestial coordinates for
  increasing values of the CR energy. In the left panels only the
  regular galactic magnetic field is included while in the right panel
  a random field component with $B_{rms}=3~\mu G$ is included. The
  dashed line represents the galactic plane with the black dot in it
  corresponding to the direct neutron flux from the GC. Darker points
  correspond to greater density in logarithmic scale.}
\label{maps}
\end{figure}

The left panels in Figure \ref{maps} show the arrival direction
distribution obtained in this way
when only the regular magnetic field  component is included and for different
values of the energy between
$10^{17.9}$ eV and $10^{18.3}$ eV.
The most intense flux arrives from the GC direction
($\alpha_{GC}=266.4^\circ$ and $\delta_{GC}=-28.9^\circ$) 
and corresponds to
the direct neutrons. A thin (nearly horizontal) strip extending from
the GC 
towards the north and south galactic poles (i.e. perpendicular to the galactic
plane, shown as a dashed line)
is due to neutrons decaying near the Earth in our
spiral arm and in the nearest inner arm respectively. The more extended excess
appearing for right ascensions between $300^\circ$ and $360^\circ$
correspond to neutrons that decayed in the inner galactic region, with the
protons reaching the Earth through the spiral arm. 
We can see that the region of the sky where events are spread
shrinks considerably with increasing energy, in
particular, at $10^{18.3}$ eV almost all the events
arrive outside the field of vision of AGASA ($\delta> - 24^\circ$).
Notice that the flux intensity of the plots with different energy
should not be compared at this point, as the energy spectrum of the
source has not yet been introduced. 

\begin{figure} [!ht]
\begin{center}
\includegraphics[width=12cm]{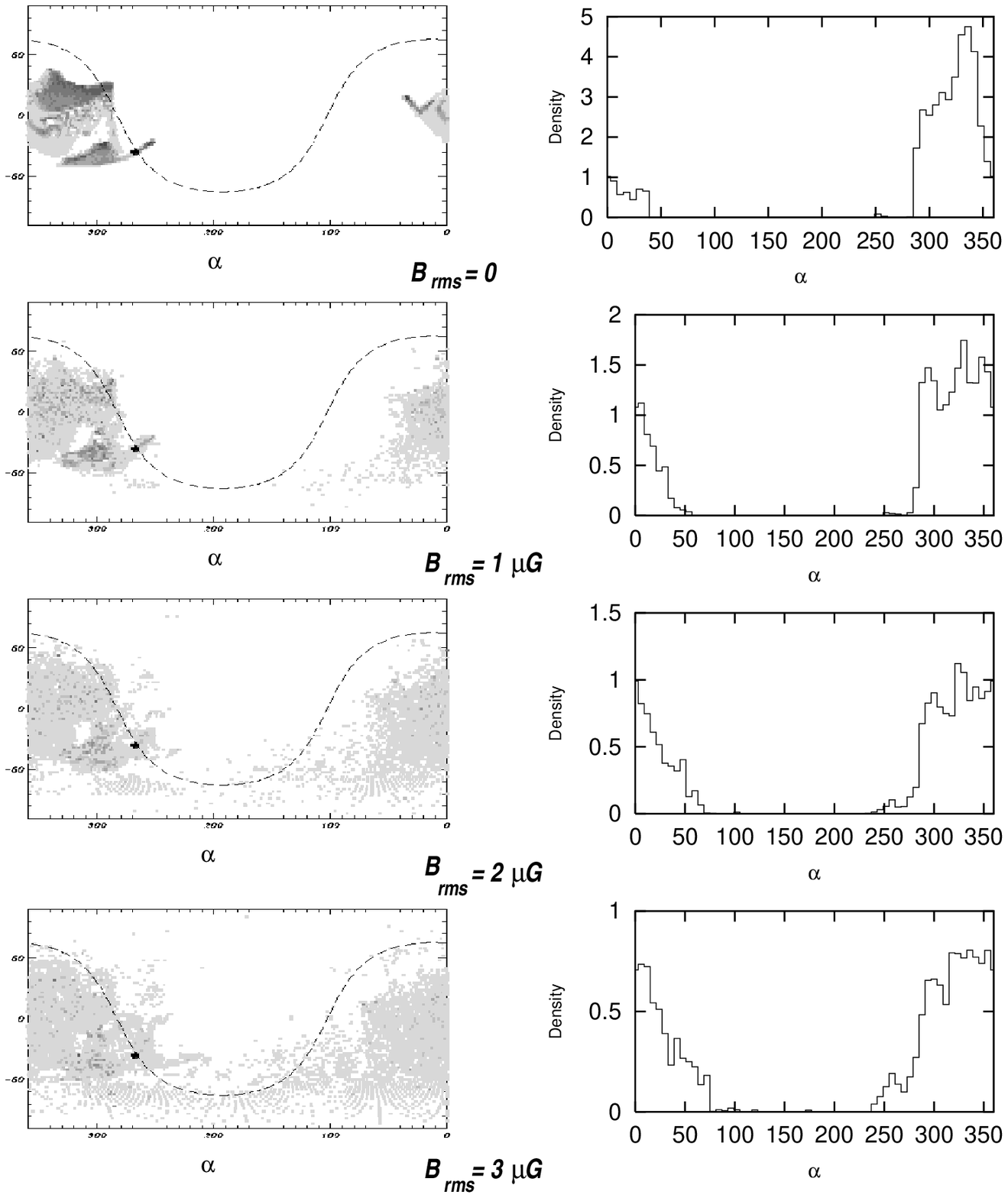} 
\end{center}
\caption
{The left panels show the arrival direction distribution in celestial 
coordinates
for different values of turbulent field strength $B_{rms}=
0, 1, 2$ and $3~\mu$G. In all plots the energy is $E = 1$ EeV.
The right panels show the corresponding right ascension distribution
taking into account the AGASA exposure.}
 \label{con random}
\end{figure}

The inclusion of the turbulent magnetic field component leads to a
spreading in the arrival direction distribution as illustrated in the
right panels of Figure \ref{maps}, in which $B_{rms}=3~\mu$G was adopted.
The spread in the arrival directions increases with the strength of
the random component, as indicated by Equation~(\ref{deflection}), and this
is clearly seen in the left panels of Figure \ref{con random}, which show
this distribution for a fixed energy of $10^{18}$~eV and increasing
intensity of the random field   ($B_{rms}=0,1,2$ and $3~\mu$G).

\begin{figure} [!ht]
\begin{center}
\includegraphics[width=12cm]{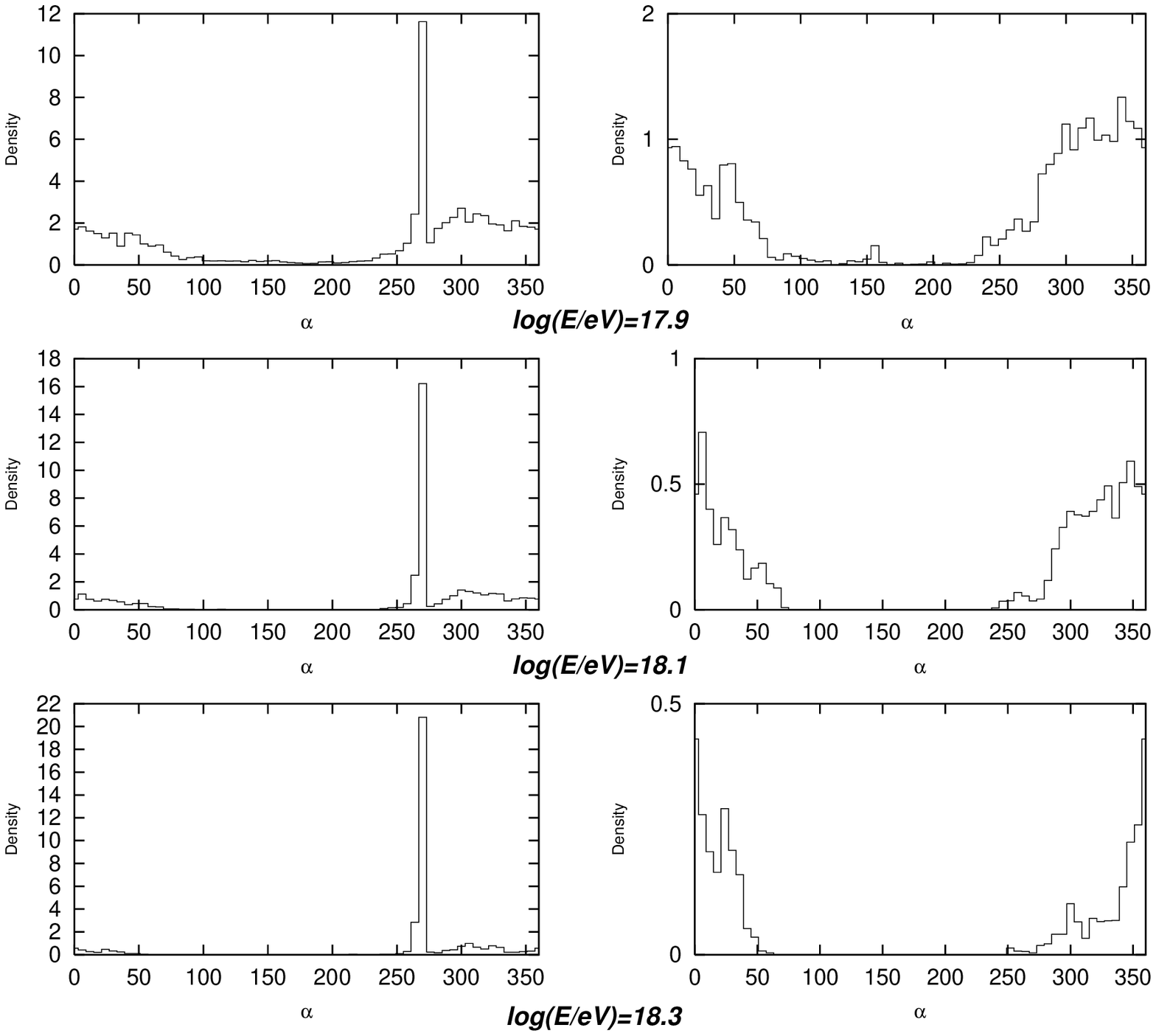} 
\end{center}
\caption
{ Right ascension density distribution corresponding to the a random
  field with $B_{rms}=3~\mu G$ for different values of the CR energy.
The left panels are for the full sky maps, while in the right panels
  the exposure of AGASA is used.
The scale is in arbitrary units.} \label{rab3}
\end{figure}

The right ascension density distribution for the $B_{rms}=3~\mu$G
random field maps of Figure \ref{maps} is shown in the left panels of
Figure \ref{rab3}. 
The sharp peak in $\alpha \sim 265^\circ$ is mainly due
to the direct neutrons. This is relatively more important
as the energy is increased, due to the increase in the neutron's decay
length. The broader excess associated to the secondary
protons arriving along the spiral arm becomes strongly suppressed above
$10^{18}$~eV due to the increasing difficulty  for the regular field to
confine the CRs along the spiral arms.  
This suppression is clearly expected to be more important as
the value of $B_{rms}$ becomes stronger, 
while on the other hand a larger value of the
regular component $B_0$ should lead to a signal more concentrated in
the spiral arm direction and surviving up to larger energies.

In order to compare our results with AGASA's measurements, it is necessary
to take into account the \emph{exposure} ($\omega$)  of the
observatory, which
gives the time-integrated effective collecting area from each
direction of the celestial sphere. 
The arrival direction distribution measured by an
observatory  should be proportional to the actual distribution multiplied 
by $\omega$.

The exposure is expected to be homogeneous in right ascension ($\alpha$) 
and to have a declination ($\delta$) dependence given by \cite{exposure}
\begin{equation} \label{expo rel}
  \omega(\delta) \propto \cos(a_0)\cos(\delta)\sin(\alpha_m)\ +
  \alpha_m\sin(a_0)\sin(\delta),
\end{equation}
where $\alpha_m$ is given by
\begin{eqnarray}
   \alpha_m = \left\{ \begin{array}{ll}
    0               & \textrm{if $\xi >  1$} \\
    \pi             & \textrm{if $\xi < -1$} \\
    \cos^{-1}(\xi)  & \textrm{otherwise}
   \end{array} \right.  \nonumber
\end{eqnarray}
and
\begin{eqnarray}
   \xi \equiv \frac{\cos(\theta_m) - \sin(a_0)\sin(\delta)}
   {\cos(a_0)\cos(\delta)}, \nonumber
\end{eqnarray}
where $a_0$ is the detector latitude  (in the case of AGASA
$a_0 = 35.78^\circ$) and $\theta_m$ is the maximum zenith angle observed.
The analysis of AGASA data correspond to events with zenith angle
$\theta < 60^\circ$. 

Taking into account the exposure, the right ascension distribution
expected to be measured at the AGASA latitude looks as shown in the
right panels of Figure \ref{rab3} for a random field amplitude
$B_{rms}=3~\mu$G.
The main difference with the full sky results displayed in the left
hand panels is the disappearance of the point-like density
peak due to the direct neutrons.
The right panels of Figure~\ref{con random} also show this
distribution for different values of $B_{rms}$ and for $E=10^{18}$eV,
for the AGASA exposure. The increasing spread of the signal associated
to the spiral arm, which is the dominating one in this case, is
apparent.  

Anisotropies in the CR distribution are often measured performing an
harmonic analysis of the right ascension distribution of the events
\cite{linsley}. This takes advantage of the uniform exposure in right
ascension of most experiments.
The right ascension distribution is related to the differential
spectrum of CR, ${\rm d}J/{\rm d}\Omega (E)$ through
$$F(\alpha)=\int {\rm d}\delta \cos \delta \omega(\delta) 
\frac{{\rm d}J}{{\rm d}\Omega}.$$
Expanding it as $F(\alpha)=F_0+F_1\cos(\alpha-\alpha_1)+\dots,$
the amplitude of the anisotropy is defined as the ratio of the dipole
to the monopole intensities: $\Delta\equiv F_1/F_0$.
 In Table \ref{resultados four}, the phase of the first harmonic
($\alpha_1$) 
\begin{table} [!ht]
\caption {Phases of the first harmonic from 
the analysis of the right ascension distribution
as a function of energy and turbulent magnetic field strength, for the
exposure at the AGASA latitude and at $35^\circ$ south.}
\begin{center}
\begin{indented}
\item[]\begin{tabular}{@{}l l l l}
\br
  $B_{rms}$& log(E) & AGASA & SOUTH  \\
\mr
  0 $\mu$G & \    & 328 & 302 \\
  1 $\mu$G & 18.0 & 333 & 303 \\
  2 $\mu$G & \    & 339 & 301 \\
  3 $\mu$G & \    & 341 & 298 \\
\mr
  \        & 17.9 & 340 & 308 \\
  3 $\mu$G & 18.1 & 344 & 291 \\
  \        & 18.2 & 354 & 286 \\
  \        & 18.3 & 358 & 279 \\
\br
\end{tabular}
\end{indented}
\label{resultados four}
\end{center}
\end{table}
from the analysis of the right ascension distribution
are listed as a function of energy and turbulent magnetic
field strength, taking into account AGASA's
exposure, and also for a hypotetical detector in the southern 
hemisphere (at a latitude $-35^\circ$, similar to the one of AUGER or
Sidney).

\subsection{Amplification}
\label{ampli}

\begin{figure} [!ht]
\begin{center}
\includegraphics[width=8cm]{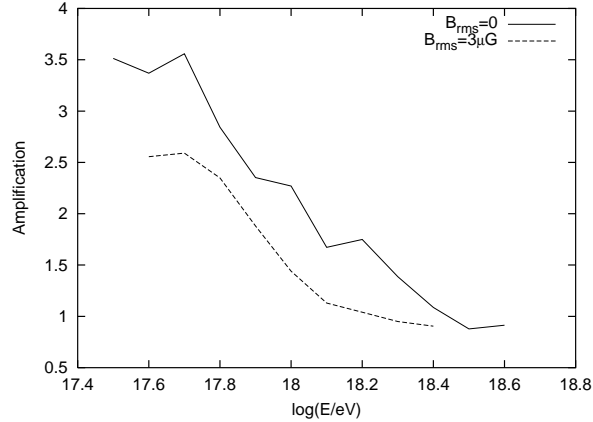} 
\end{center}
\caption{Amplification vs. energy of a neutron source located at
the GC due to the GMF, with (solid line) and without (dotted line)
including the turbulent component of the GMF ($B_{rms}=3\ \mu$G)
in the simulation.}
\label{amplificacion}
\end{figure}

The total flux that we receive from a given source  is modified by the
presence of a magnetic field. This can be described by introducing 
the amplification,
$A(E)$, which is the ratio of the flux of particles arriving to the
Earth from any direction in the presence of the GMF, $J(E)$, 
and that in the absence of
magnetic field, $J_0(E)$, i. e. $A(E)=J(E)/J_0(E)$. The amplification
is a function of the energy and the mean effect turns out to be larger 
as the energy decreases, as expected since the CRs tend to be trapped 
more efficiently in the spiral arms of the GMF. At larger energies
the amplification tends to unity, as protons are less affected by the
magnetic field and a larger fraction of the neutrons arrive to the Earth 
before decaying.
Figure \ref{amplificacion} shows the amplification vs. energy
for adopted values of the random magnetic field of $B_{rms}= 0$ and 
$3~\mu$G. The smaller amplification obtained in the presence of the
random magnetic field is due to the enhancement of the escape of CRs
from the spiral arms produced by this component.

\subsection{Anisotropy}

The anisotropy detected by AGASA arises in a narrow range of energies from
$E=10^{17.9}$eV to $E=10^{18.3}$eV. It has been suggested that the 
lack of anisotropies at
smaller energies could be explained in the model discussed here
because of the shorter decay length of the neutrons and the deflection
of the resulting protons in the galactic magnetic field. In order to
test quantitatively this hypothesis we have to estimate the total
anisotropy produced by the GC neutron source in the presence of the
background of all the other sources of CR in that energy range.

The flux of CRs with energy around 1~EeV arriving to the Earth is
thought to come from galactic sources, such as supernovae, that give
the main  contribution below the ankle ($E_{ankle}\simeq 5\times
10^{18}$~eV), as well as extragalactic sources, which dominate above
the ankle. This can be splitted into the flux from the GC source,
${\rm d}J/{\rm d}\Omega (E)$, and the contribution from all the other 
sources, ${\rm d}J_{bkg}/{\rm d}\Omega (E)$, i.e.
${\rm d}J_{tot}/{\rm d}\Omega ={\rm d}J_{bkg}/{\rm d}\Omega
+{\rm d}J/{\rm d}\Omega $. 

The total (measured) differential energy spectrum of CRs can be
approximated by \cite{na00}
\begin{equation} \label{flujo}
 \frac{{\rm d}J_{tot}}{{\rm d}\Omega}(E) = C \times 
\bigg( \frac{E}{6.3 \times 10^{18}} \bigg)^{-3.2 \pm 0.05}
\end{equation}
in the range of energies from 
$4\times 10^{17}\ \uni{eV}< E < 6.3\times 10^{18}\ \uni{eV}$,
and $C = (9.23 \pm 0.65)\times 10^{-33} \uni{m}^{-2} \uni{s}^{-1} 
\uni{sr}^{-1} \uni{eV}^{-1}$.

The intensity of the total flux received at the Earth from the source
is $J(E)=J_0(E)A(E)$, where
$A(E)$ is the amplification computed in the previous section and for the
actual source flux we can consider a power law dependence with energy
$J_0(E)\propto E^{-2.2}$, which is typical of Fermi acceleration in shocks.

\begin{figure} [!ht]
\begin{center}
\includegraphics[width=8cm]{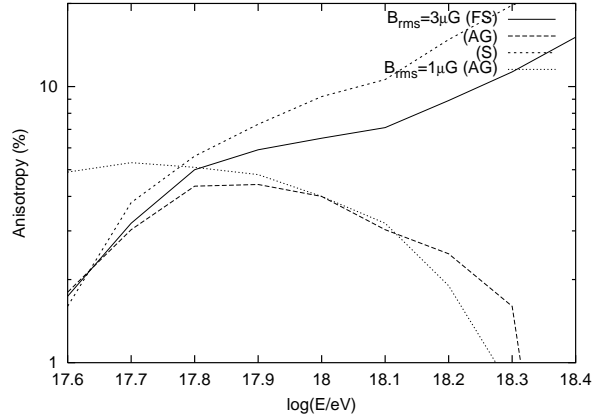} 
\end{center}
\caption
{Anisotropy as a function of energy for random magnetic fields of
  strengths $B_{rms}=1$ and $3~\mu$G, and for exposures corresponding
  to full sky (FS), AGASA (AG) and for a southern observatory at 
$b=-35^{\circ}$ (S).}
 \label{aniso}
\end{figure}

If we assume that the main contribution to the anisotropy comes from
the GC neutron source, and hence consider that the
background CR flux is essentially isotropic, we can write the observed
anisotropy as $\Delta(E)=\Delta_s(E) f(E)$, where 
$f(E)\equiv F_{0(s)}/F_{0(bkg)}$ is the ratio of the contributions to
the right ascension monopole from the source flux and the background
flux. For an anisotropic flux, as in this case, $f(E)$ depends on the
exposure of the experiment.
The anisotropy of the source flux, $\Delta_s(E)$, 
can be computed from the harmonic analysis of the right ascension
distribution of the GC events as discussed in Section \ref{add}. 
We show in Figure  \ref{aniso} the
expected anisotropy as a function of the energy for different values
of the turbulent magnetic field  amplitude and exposures.
For a random field amplitude $B_{rms}=3~\mu$G, the anisotropy
grows sharply around $E=10^{17.8}$ as observed by AGASA. 
This is not the case
for smaller values of the rms amplitude of the turbulent field as
is apparent from the curve labeled $B_{rms}=1~\mu$G in the Figure, 
as the CR arrival directions are not spread sufficiently at small
energies to wash out the anisotropies. 
Notice that the effects associated to a harder
GC source spectrum are partially compensated by the effects due to the
increased amplification  of the fluxes for decreasing energies,
leading to a more moderate energy dependence. 
We have normalized the source flux in
such a way that the anisotropy is 4$\%$ at $E\simeq10^{18}$eV for the 
AGASA exposure. This corresponds to a fraction $f\simeq 0.026$ 
contributed by the GC source to the right ascension monopole at that
energy. 

 The sharp fall in the anisotropy appearing at $E\simeq 10^{18.3}$~eV
 for the AGASA curve arises
 because above that energy most CRs arrive as direct neutrons from the GC
 direction, which is out of the field of view of AGASA. 
 The possible existence of a cutoff in the acceleration energy of the source 
around that range of values is actually not necessary to explain AGASA's
 results.

\subsection{Source luminosity}

We can estimate the luminosity of the GC source which corresponds to the
normalization of its flux at the Earth adopted in the previous Section
to fit the
4$\%$ amplitude of the right ascension first harmonic measured by AGASA.
This is given by
\begin{equation}
  L_{GC} = 4\pi r_0^2 \int \rd E E J_0(E),
\end{equation}
where $r_0$ is the distance to the GC, and the source flux in the 
absence of magnetic fields is given by $J_0(E)=J(E)/A(E)$.
The source flux $J(E)$ at $10^{18}$eV can be related to the background
flux $J_{bkg}$ at that energy using the value of $f(E)$
obtained in the last Section. With the
amplification computed in Section \ref{ampli}, the local flux at
$10^{18}$eV results 
$J_0(10^{18}~{\rm eV})\simeq 1.3\times10^{-30}m^{-2}s^{-1}{\rm eV}^{-1}$.
For a power law energy spectrum of the source $J_0(E)\propto
E^{-2.2}$, the total luminosity in the decade of energy from
$10^{17.5}$ to $10^{18.5}$ then results
$L_{GC}(10^{17.5}-10^{18.5})\simeq 4\times 10^{36}$~erg/s. 
This value is more than an order of magnitude lower than the 
maximum luminosity
($10^{38}$ erg/s) estimated for Sgr A* \cite{Sgr A} mentioned in the
Introduction. 

The expected flux of direct neutrons is a factor 
$e^{-r_0/\gamma c \tau_n}$ of the source flux $J_0(E)$.
The integral of this
in the energy range from $10^{17.9}$~eV to $10^{18.5}$~eV 
corresponds then to a direct neutron flux of
$2\times 10^{-13}{\rm m}^{-2}s^{-1}$, which is not very different 
from the value  $(9\pm3)\times 10^{-14}{\rm m}^{-2}{\rm s}^{-1}$ 
reported by SUGAR for their point-like excess \cite{sugar}.
 
\begin{figure}
\begin{center}
\includegraphics[width=9cm]{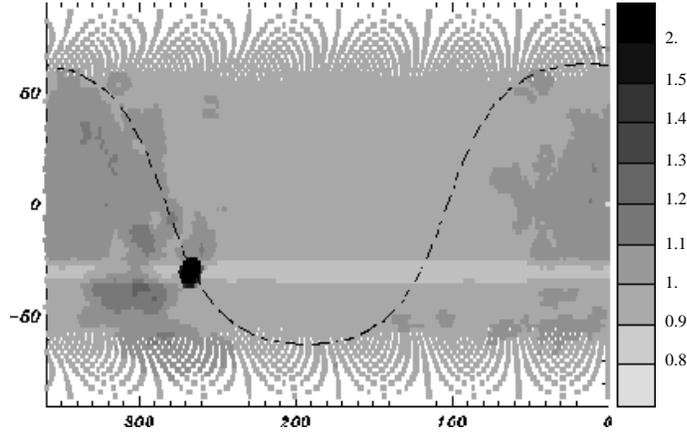} 
\end{center}
\caption { Ratio of the density in each direction to the mean density
  at that declination for $E=1$ EeV and $B_{rms}=3\ \mu$G. }
\label{excessmap}
\end{figure}

In order to give an idea of the distribution of the excess of events 
over the ones expected for an isotropic distribution, we
made a map (similar to those constructed by AGASA and SUGAR) 
in which we added to an isotropic background the CR density
expected from a GC neutron source of the required  intensity to fit
the AGASA's right ascension anisotropy. 
The map in Figure
\ref{excessmap} displays the ratio of the intensity at each direction 
to the mean density at that declination (following AGASA's analysis). 
We have smoothed the map  with a
window of $3^{\circ}$ radius to take into account the typical angular
resolution at these energies.
Although the
particular details of the map will depend on the GMF structure and
strength, that are poorly known, we think that the main characteristic
of the distribution are picked with the realistic model adopted. 
Notice however that the proton signal around the GC is sensitive to
the actual location of the reversals in direction of the regular
magnetic field. In the model adopted, in which the nearest reversal
to the GC direction is at less than 1~kpc from the Earth, the excess
observed at southern galactic latitudes is more significant than the
one at northern latitudes, but this could change if the reversal was
at a smaller galactocentric radius. Also, a vertical component of the
magnetic field could displace these excesses in the direction parallel to
the  galactic plane.
 
\section{Conclusions}

We examined the possibility that a source of neutrons at the GC
could be responsible for the anisotropy measured by AGASA and SUGAR.
As the GC itself is out of the field of view of AGASA, this
observatory would not detect direct neutrons from a source in the GC,
but it would just see the protons produced by the  neutron decays. As their
trajectories are bent by the GMF, the distribution of protons arriving
to the Earth depends strongly on the energy of the particles and on
the structure and strength of the field, which are not well known. We
assumed a particular model for the regular field, which is consistent
with the main observational evidences, and considered different
reasonable amplitudes for the turbulent magnetic field.
We found that there is a sharp rise in the anisotropy at energies
around 1~EeV, as observed by AGASA, for the largest amplitude of the 
turbulent magnetic field considered, $B_{rms}=3~\mu$G. At energies
larger than 2~EeV, essentially no particles are expected from the GC source
in the field of view of AGASA, what explains the disappearance of the
anisotropy independently of the source acceleration cutoff. The
predictions are however very different for an observatory in the
south. This one should detect the strong point-like signal from the direct
neutrons in addition to the extended signal from the protons. At
energies below 1~EeV the neutron signal is suppressed due to the
reduced decay length of the neutrons and the expected anisotropy has a
similar behavior as that of a northern observatory. At larger
energies, and up to the maximum acceleration energy of the source, the
point-like signal should be clearly seen. It is interesting to notice
that the neutron source luminosity needed to fit the $4\%$ right
ascension anisotropy detected by AGASA (due to the protons in this
model) give rise to  a point-like neutron flux consistent with the one
measured by SUGAR. However, unless there is some systematic pointing
error in their data, the offset of the signal from the GC direction
cannot be explained. Moreover, for the GMF adopted, the phase of the
first harmonic in right ascension obtained ($\sim 330^\circ$) is
somewhat larger than AGASA's one.
Another important point is that within this scenario the
excess observed along the Cygnus region is also related to the central
source (for this it is crucial that the regular magnetic field be
along the spiral arm and not just azimuthal). This excess disappears
for $E>2\times 10^{18}$~eV due to the impossibility for the GMF to
confine the CRs along the spiral arms, independently of the source
upper energy cutoff. 
The alternative explanation of a diffusive  flux of charged
particles from the GC direction \cite{te01,ca02} should give rise to a
wide angle anisotropy, with a deficit of particles in the galactic
anticenter direction and a cosine angular dependence with respect to 
the direction of maximum flux, quite different from the neutron source
signal. 
More data from an observatory in the south would be very helpful to 
confirm or reject these hypothesis.
Although the AUGER observatory \cite{auger} in construction in
Argentina is focused to the study of CRs with energy above
$10^{19}$~eV, 
it may be able to detect lower energy events (down to $10^{18}$~eV).
A dedicated detector for the $10^{17}$~eV to $10^{19}$~eV energy range
is also under study \cite{ad03}.

If the sharp peak in the GC direction is seen, confirming
the neutron source hypothesis, then the
distribution of the protons would give information about the
structure of the GMF.

\ack
Work supported by CONICET and Fundaci\'on Antorchas.

\end{document}